\pdfoutput=1

\documentclass[11pt]{article}

\usepackage{acl}

\usepackage{times}
\usepackage{latexsym}
\usepackage{subfigure}
\usepackage{graphicx}
\usepackage{caption}

\usepackage[T1]{fontenc}

\usepackage[utf8]{inputenc}

\usepackage{microtype}

\title{hate-alert@DravidianLangTech-ACL2022: Ensembling Multi-Modalities for Tamil TrollMeme Classification}

\author{Mithun Das, Somnath Banerjee, Animesh Mukherjee \\
  Indian Institute of Technology, Kharagpur, India \\
  mithundas@iitkgp.ac.in, som.iitkgpcse@kgpian.iitkgp.ac.in, animeshm@cse.iitkgp.ac.in \\}

\begin{document}
\maketitle
\begin{abstract}
Social media platforms often act as breeding grounds for various forms of trolling or malicious content targeting users or communities. One way of trolling users is by creating memes, which in most cases unites an image with a short piece of text embedded on top of it. The situation is more complex for multilingual(e.g., Tamil) memes due to the lack of benchmark datasets and models. We explore several models to detect Troll memes in Tamil based on the shared task, "Troll Meme Classification in DravidianLangTech2022" at ACL-2022. We observe while the text-based model MURIL performs better for Non-troll meme classification, the image-based model VGG16 performs better for Troll-meme classification. Further fusing these two modalities help us achieve stable outcomes in both classes. Our fusion model achieved a 0.561 weighted average F1 score and ranked \textbf{second} in this task.
\end{abstract}

\section{Introduction}

Over the past few years, social media platforms have been expanding rapidly. Users of the platform interact by sharing content to enrich their knowledge and social connections. Although most of the content on social media platforms that existed so far was textual, recently, a unique message was born: the \textit{meme}. A meme is usually created by an image and a short piece of text on top of it, entrenched as part of the image. Memes are generally meant to be harmless and conceived to look humorous, but sometimes, bad actors use memes for threatening and abusing individuals or specific target communities. Such memes are collectively known as Offensive/Troll memes in social media. 

Trolling is the exercise of publicizing a message via social media that is planned to be abusive, inciting, or threatening to distract, which often has rambling or off-topic content to provoke the audience\cite{bishop2014dealing,suryawanshi2020multimodal}. In addition, such memes can be treacherous as they can easily harm the reputation of individuals, famous celebs, political entities, businesses, or social groups, e.g., minorities. Although various studies have been conducted to detect offensive posts using different natural language techniques, Troll meme classification has not yet been explored.
 
The situation for countries like India is more complicated due to the immense lanuage diverisy\footnote{\url{https://en.wikipedia.org/wiki/Languages_of_India}}.  The meme in the Indian context, can be composed in English, local language (native or foreign script) or in combination of both language and script. This adds another challenge for the troll meme classification.

\begin{figure}[t!]
\centering
\begin{minipage}{.4\textwidth}
    \centering
        \centering
        \includegraphics[scale=0.3]{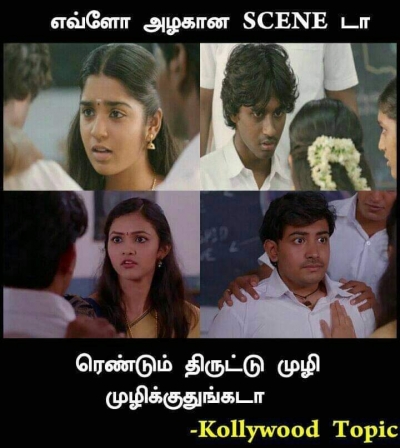}
        \caption*{(a) An example of a Troll meme}
        \end{minipage}
    \quad
    \begin{minipage}{.4\textwidth}
        \centering
        \includegraphics[scale=0.5]{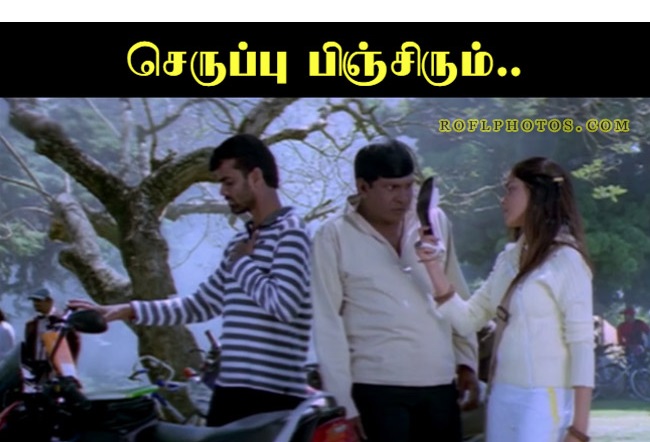}
        \caption*{(b) An example of a Non-troll meme}
    \end{minipage}
    \caption{ Examples of troll and not-troll meme}
    \label{fig:ExampleImage}
\end{figure}

Recently, there has been a lot of effort to investigate the malicious side of memes, e.g., focusing on hate\cite{gomez2020exploring}, offensive\cite{suryawanshi2020multimodal}, and harmful\cite{pramanick2021momenta} memes. However, the majority of the studies are centralized around the English language. Further several shared tasks like HASOC 2021\cite{modha2021overview}, DravidianLangTech 2021\cite{chakravarthi2021findings}, have been organized on multiple languages for hostile content detection in the Indian context, but it is limited to textual classification. Extending those tasks further, the organizer of this shared task has organized a classification task to identify troll memes in Tamil by providing 2,967 memes. This paper illustrates the methodologies we used to identify Tamil troll memes, which helped us achieve \textbf{second} place in the final leader-board standings of shared tasks.

\begin{table}
\centering
\begin{tabular}{|l|l|l|l|}
\hline
\textbf{Split} & \textbf{Troll} & \textbf{Non-troll} & \textbf{Total} \\ \hline
\textbf{Train} & 1,282 & 1,018 & 2,300 \\ \hline
\textbf{Test} & 395 & 272 & 667 \\ \hline
\textbf{Total} & 1,677 & 1,290 & 2,967 \\ \hline
\end{tabular}
\caption{Dataset statistics}
\label{tab:dataset}
\end{table}

\begin{figure}
  \centering
	\includegraphics[scale=0.6]{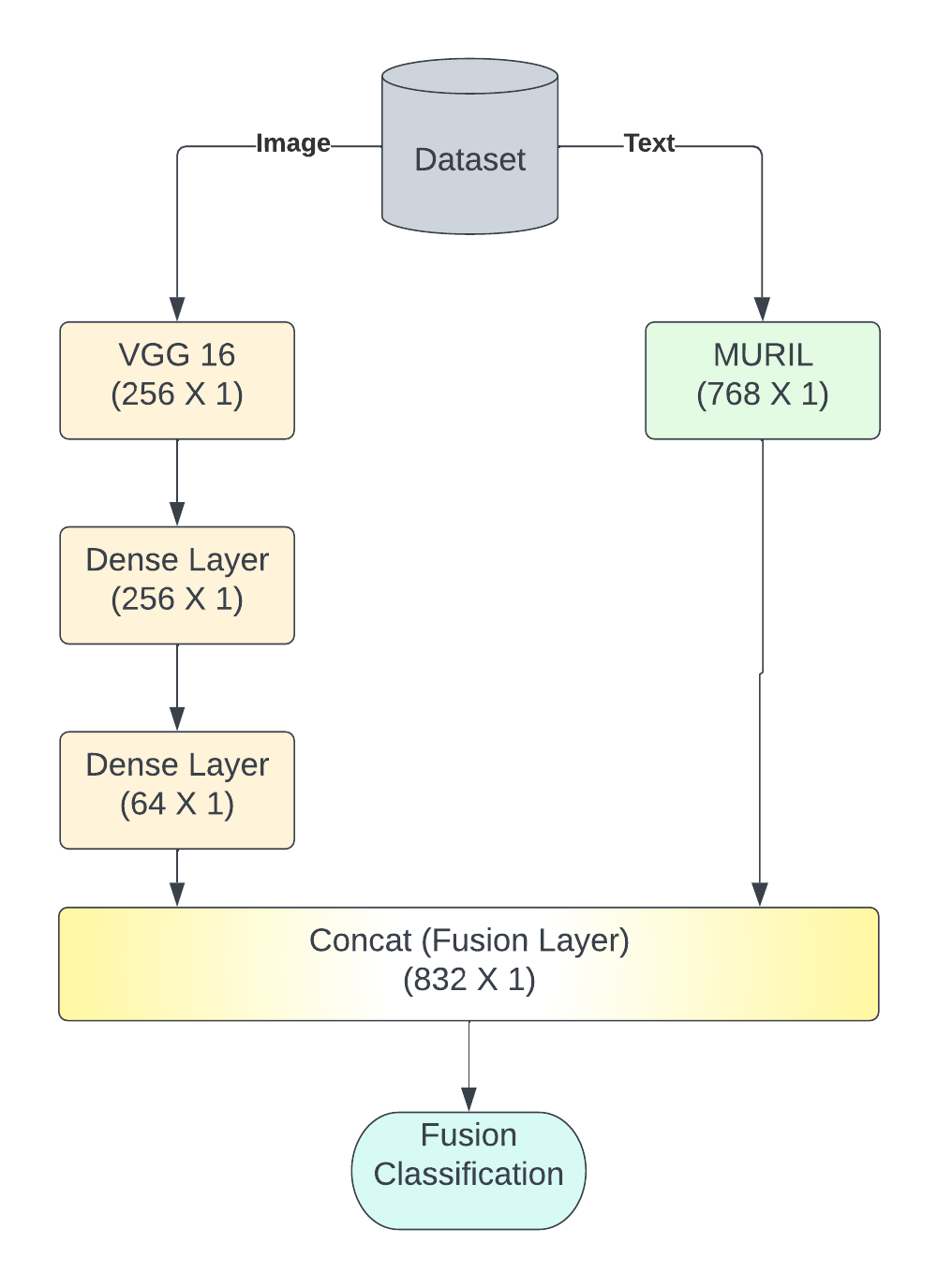}
    \caption{Our fusion model architecture with VGG16 and MURIL}
	\label{fig:fusion_model}
\end{figure}

\begin{table*}
\centering
\begin{tabular}{|l|l|l|l|l|l|}
\hline
\textbf{Model} & \textbf{Accuracy} & \textbf{F1 Score(T)} & \textbf{F1 Score(w)} & \textbf{Precision(w)} & \textbf{Recall(w)} \\ \hline
\textbf{MURIL} & 0.556 & 0.637 & {\underline{0.552}} & {\underline{0.549}} & 0.556 \\ \hline
\textbf{VGG16} & \textbf{0.587} & \textbf{0.736} & 0.458 & 0.522 & \textbf{0.587} \\ \hline
\textbf{Fusion} & {\underline{0.566}} & {\underline{0.649}} & \textbf{0.561} & \textbf{0.558} & {\underline{0.567}} \\ \hline
\end{tabular}
\caption{Performance Comparisons of Each Model. T: Troll Class. w: Weighted-Average. The best performance in each column is marked in \textbf{bold} and second best is \underline{underlined}}
\label{tab:results}
\end{table*}

\begin{figure*}
  \centering
	\includegraphics[scale=0.4]{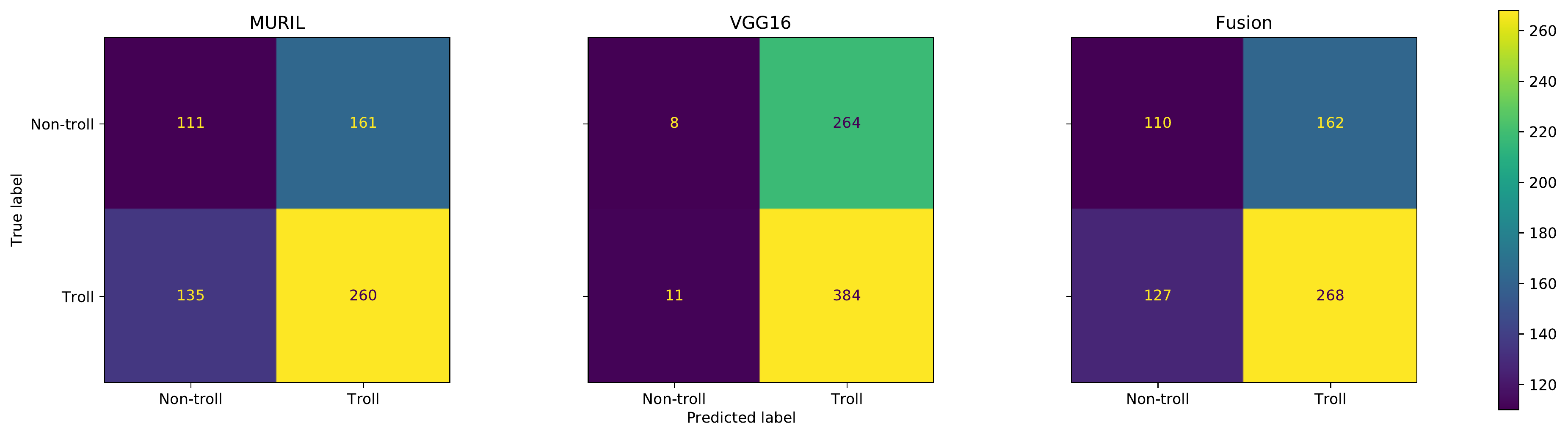}
    \caption{Confusion Matrix on Test Data for Each Model}
	\label{fig:confusion_matrix}
\end{figure*}

\section{Related Work}

This section discusses some of the text-based abusive content detection methods and briefly explains the multi-modal techniques used so far to detect malicious memes.

\subsection{Text-based abusive content detection}
Recently, a lot of work has been carried out to identify abusive speech using text from social media posts~\cite{das2020hate}. In 2017, Davidson et al. ~\shortcite{Davidson2017AutomatedHS} made public a Twitter dataset in which thousands of tweets were labeled offensive, hate, and neither. The earlier efforts to create such classifiers used easy methods such as linguistic features, word n-grams, bag-of-words, etc~\cite{Davidson2017AutomatedHS}. With the availability of larger datasets, researchers have started utilizing complex models such as deep learning and graph embedding\cite{das2021you} strategies to improve the classifier performance of hate speech detection in social media posts. In 2018, Pitsilis et al.~\shortcite{Pitsilis2018DetectingOL} used deep learning-based models, such as the recurrent neural networks (RNNs), to detect the abusive tweets in the English language and witnessed that it was pretty effective in this task. In contrast, RNNs have been established to perform well with several language models. In addition, other neural network models, such as LSTM and CNN, have succeeded in detecting abusive speech~\cite{Goldberg2015,Sarracn2018HateSD}. Recently, Transformer-based~\cite {Vaswani2017AttentionIA} language models such as BERT,~\cite{Devlin2019BERTPO} are becoming quite prevalent in several downstream tasks, such as spam detection, classification\cite{das2021abusive,banerjee2021exploring}, etc. Having observed the exceptional performance of these Transformer based models, we also utilize a Transformer based model, MURIL, which is pre-trained explicitly in Indian Languages.

\subsection{Multi-modal abusive content detection} 
Lately, several datasets have been made public to the research community for abusive meme detection. Sabat et al.\shortcite{sabat2019hate} created a dataset of 5,020 memes for hate speech detection. The MMHS150K hate meme dataset developed by  Gomez et al.\shortcite{gomez2020exploring} is one of the enormous datasets collected from Twitter, consisting of 150K posts. Similarly, Facebook AI~\cite{kiela2020hateful} introduced another Hateful Meme dataset of 10K+ posts labeled hateful and non-hateful. As part of the hateful meme detection, an array of techniques with diverse architecture ranging from the text-based model, image-based model, and multi-modal models have been employed, including Glove embedding, FastText embedding, ResNet-152, VGG16, VisualBERT, UNITER, ViLBERT CC, V-BERT COCO\cite{pramanick2021momenta,chandra2021subverting}.

In this work, we use the VGG16 model, which is extensively used for several classification problems, to extract the features of all the memes and finally use it with the textual features to design our final model.

\section{Dataset Description}

The shared task on Troll Meme Classification in DravidianLangTech2022~\cite{trollmeme-eacl} at ACL-2022  is based on a classification problem with the aim of moderating and minimizing the offensive/harmful content in social media. The objective of the shared task is to devise methodologies and vision-language models for troll meme detection in Tamil. We show the class distribution of the dataset\cite{suryawanshi-etal-2020-dataset,dravidiantrollmeme-eacl} in Table \ref{tab:dataset}. The training set consisting of 2,300 memes (out of which 1,282 memes were labeled as troll meme) and  the test set consisting of 667 memes. In addition, the latin transcribed texts were shared for all memes. We show example of both Troll and Non-troll memes in Figure \ref{fig:ExampleImage}.

\section{Methodology}
In this section, we discuss the different parts of the pipeline that we pursued for the detection of troll meme using the dataset.

\subsection{Uni-modal Models}
As part of our initial experiments, we created the following two uni-model models, one utilizing text features and the other using image-based features.

\noindent\textbf{MURIL:} MURIL\cite{khanuja2021MURIL} is a transformer encoder having 12 layers with 12 attention heads and 768 dimensions. We used the pre-trained model which has been trained on 17 Indian languages and their transliterated counterparts using the MLM (masked language model) and the next sentence prediction (NSP) loss functions. The dataset used for pre-training is obtained by using the publicly available corpora from Wikipedia and Common Crawl. We pass all the texts associated with the meme via MURIL to get the 768-dimensional feature vectors for each meme and then finally fed it to a output node for the final prediction.

\noindent\textbf{VGG16:} VGG16~\cite{simonyan2014very} is a Convolutional Neural Network architecture,  a variant of the VGG model which consists of 16 layers and is very appealing because of its very uniform architecture. We pass all the images(meme) via VGG16 and get the 256-dimensional feature vectors, then we pass it to the two dense layer of size 256 (with dropout of 0.5), 64 and finally fed it two the output node for the final prediction.

\subsection{Fusion Model}
The uni-modal models we used so far do not use the relation between the text and image present in the meme. To have better understanding between the text and image, we design a new MURIL+VGG16 fusion classifier, where we first concatenate the embedding from the both MURIL and VGG16 models discussed above, then we pass the concatenated embedding to a classification node for the final prediction. The detail of the pipeline is presented in Figure \ref{fig:fusion_model}.

All the models are trained with binary cross-entropy loss functions and Adam optimizer for 20 epochs.

\section{Results}
Table \ref{tab:results} demonstrates the performance of each model. We observe among the uni-modal models, VGG16 has the highest Accuracy(MURIL: 0.556, VGG16: 0.587) and F1 score (MURIL: 0.637, VGG16: 0.736) for troll class. Though in terms of weighted F1 score(MURIL: 0.552, VGG16: 0.458), the text-based model MURIL performs better. When we fuse these two models, the fusion model achieves the highest weighted F1 score(0.561) among all the models. To further understand the model's weakness, we show the confusion matrix of each model in Figure \ref{fig:confusion_matrix}. We observe that while the MURIL performs better on the Non-troll meme datapoints, VGG16 performs better on the troll meme datapoints. Whereas on the non-troll meme data points, VGG16 shows inferior performance. The fusion model brings the positive characteristics of both MURIL and VGG16 and performs the best by understanding better connections between the text and image of the memes.

\section{Conclusion}
In this shared task, we deal with a novel problem of detecting Tamil troll memes. We evaluated different uni-modal models and introduced a fusion model. We found that text-based model MURIL performs better on the Non-troll class, whereas VGG16 performs better on the Troll class. Ensembling these two models help us in gaining stable outcomes in both classes. We plan to explore further other vision-based models to improve classification performance as an immediate next step. 

\bibliography{anthology,custom}
\bibliographystyle{acl_natbib}


\end{document}